\documentclass{cimento}

\usepackage{epsfig}

\newcommand{\be}{\begin{equation}}
\newcommand{\ee}{\end{equation}}
\newcommand{\bq}{\begin{eqnarray}}
\newcommand{\eq}{\end{eqnarray}}

\newcommand{\D}{\mathrm{d}}

\newcommand{\I}{\mathrm{i}}

\def\lsim{\mathrel{\rlap{\lower4pt\hbox{\hskip1pt$\sim$}}\raise1pt\hbox{$<$}}}
\def\gsim{\mathrel{\rlap{\lower4pt\hbox{\hskip1pt$\sim$}}\raise1pt\hbox{$>$}}}
\def\Vec#1{\mathpalette{\VVec}{#1}}                  
\def\VVec#1#2{\mbox{\boldmath$#1#2$\unboldmath}}
\def\nostrocostruttino#1\over#2{\mathrel{\mathop{\kern 0pt \rlap
{\hbox{$#1$}}} \hbox{\kern-.135em $#2$}}}

\newlength{\dhatheight}

\newcommand{\STRUT}{\rule[-2.1ex]{0pt}{5.5ex}}        
\def\anti#1{\mathpalette{\@anti}{#1}#1}
\def\@anti#1#2{\sbox0{$#1#2$}
  \makebox[0pt][l]{$#1\kern.30\ht0\overline{\kern-.35\ht0\phantom{#2}}$}}

\title{Correlations between SIDIS azimuthal asymmetries in target and current fragmentation regions}
\author{\underline {A.~Kotzinian}\from{ins:t}\from{ins:y}
\ETC,
M.~Anselmino\from{ins:t},
        \atque
V.~Barone\from{ins:a}}
\instlist{\inst{ins:t} {\it Dipartimento di Fisica Teorica, Universit{\`a}
di Torino; \\
INFN, Sezione di Torino, 10125 Torino, Italy}
\inst{ins:y}{\it Yerevan Physics Institute, 2 Alikhanyan Brothers St., 375036 Yerevan, Armenia}
\inst{ins:a}{\it Di.S.T.A., Universit{\`a} del Piemonte
Orientale ``A. Avogadro''; \\
INFN, Gruppo Collegato di Alessandria,  15121 Alessandria, Italy}}

\PACSes{\PACSit{13.85.Fb}{Inelastic scattering: many-particle final states.}
\PACSit{13.87.Fh}{Fragmentation into hadrons.}
\PACSit{13.88.+e}{Polarization in interactions and scattering.}}

\begin{document}

\maketitle

\begin{abstract}
We shortly describe the leading twist formalism for spin and transverse-momentum dependent fracture functions recently developed and present results for the production of spinless hadrons in the target fragmentation region (TFR) of
SIDIS~\cite{Anselmino:2011ss}. In this case not all fracture functions can be accessed and only a Sivers-like single spin azimuthal asymmetry shows up at LO cross-section.
Then, we show~\cite{Anselmino:2011bb} that the process of double hadron production in polarized SIDIS -- with one spinless hadron produced in the current fragmentation region (CFR) and another in the TFR -- would  provide access to all 16 leading twist fracture functions. Some particular cases are presented.
\end{abstract}

\section{Introduction}
As it is becoming increasingly clear in the last decades, the study of the
three-dimensional spin-dependent partonic structure of the nucleon in SIDIS processes requires a full understanding of the hadronization process after the hard lepton-quark scattering. So far most SIDIS experiments were studied in the CFR, where an adequate theoretical formalism based on distribution and fragmentation functions has been established (see for example Ref.~\cite{Bacchetta:2006tn}). However, to avoid
misinterpretations, also the factorized approach to SIDIS description in the TFR
has to be explored. The corresponding theoretical basis -- the fracture functions formalism -- was established in Ref.~\cite{Trentadue:1993ka}
for hadron transverse momentum integrated unpolarized cross-section. Recently this approach was generalized~\cite{Anselmino:2011ss} to the spin and transverse momentum dependent case (STMD).

We consider the process (adopting the same notations as in
Ref.~\cite{Anselmino:2011bb})
\begin{equation}\label{sidis-tfr}
l(\ell,\lambda) + N(P,S) \to l(\ell') + h(P_h) + X(P_X)
\end{equation}
with the hadron $h$ produced in the TFR. We use the standard DIS notations
and in the $\gamma^*-N$ c.m. frame we define the $z$-axis along the direction
of $\Vec q$ (the virtual photon momentum) and the $x$-axis along ${\Vec \ell}_T$,
the  lepton transverse momentum. The kinematics of the produced hadron is defined by the variable $\zeta = P_h^-/P^- \simeq E_h/E$ and its transverse momentum
$\Vec P_{h\perp}$ (with magnitude $P_{h\perp}$ and azimuthal angle $\phi_h$).
Assuming TMD factorization the cross-section of the process (\ref{sidis-tfr})
can be written as
\begin{equation}\label{sidis-tfr-cs}
\frac{\D\sigma^{l(\ell,\lambda)+N(P_N,S) \to l(\ell')+h(P)+X}}
{\D x_B \, \D Q^2 \, \D \zeta \, \D^2 \Vec{P}_{h\perp} \, \D \phi_S} =
{\cal M} \otimes \frac{\D \sigma^{\ell(l,\lambda)+q(k,s) \to \ell(l')+q(k',s')}}
{\D Q^2}\,,
\end{equation}
where $\phi_S$ is the azimuthal angle of the nucleon transverse polarization.
The STMD fracture functions ${\cal M}$ has a clear probabilistic meaning:
it is the conditional probability to produce a hadron $h$ in the TFR when the
hard scattering occurs on a quark $q$ from the target nucleon $N$. The
expression of the non-coplanar polarized lepton-quark hard
scattering cross-section can be found in Ref.~\cite{Kotzinian:1994dv}.

The most general expression of the LO STMD fracture functions for unpolarized ($\mathcal{M}^{[\gamma^-]}$), longitudinally polarized ($\mathcal{M}^{[\gamma^-\gamma_5]}$) and transversely polarized ($\mathcal{M}^{[\I \, \sigma^{i -} \gamma_5]}$) quarks are introduced in the expansion of the leading twist
projections as~\cite{Anselmino:2011ss,Anselmino:2011bb}:
\begin{eqnarray}
\mathcal{M}^{[\gamma^-]} &=& \hat{u}_1
+ \frac{ {\Vec P}_{h\perp} \times  {\Vec S}_\perp}{m_h} \, \hat{u}_{1T}^h
+ \frac{ {\Vec k}_\perp \times {\Vec S}_\perp}{m_N} \, \hat{u}_{1T}^{\perp}
+ \frac{S_\parallel \, ( {\Vec k}_\perp \times  {\Vec P}_{h\perp})}{m_N \, m_h}
  \, \hat{u}_{1L}^{\perp h} \label{up-frf} \\
\mathcal{M}^{[\gamma^-\gamma_5]} & = &
S_\parallel \, \hat{l}_{1L}
+ \frac{{\Vec P}_{h\perp} \cdot {\Vec S}_\perp}{m_h} \, \hat{l}_{1T}^h
+ \frac{ {\Vec k}_\perp \cdot  {\Vec S}_\perp}{m_N} \, \hat{l}_{1T}^{\perp}
+ \frac{ {\Vec k}_\perp \times {\Vec P}_{h\perp}}{m_N \, m_h} \,
  \hat{l}_1^{\perp h} \label{lp-frf} \\
\mathcal{M}^{[\I \, \sigma^{i -} \gamma_5]} & = & S_\perp^i \, \hat{t}_{1T}
+ \frac{S_\parallel \, P_{h\perp}^i}{m_h} \, \hat{t}_{1L}^h
+ \frac{S_\parallel \, k_\perp^i}{m_N} \, \hat{t}_{1L}^{\perp}
\nonumber \\
& & + \, \frac{( {\Vec P}_{h\perp} \cdot {\Vec S}_\perp)
\, P_{h\perp}^i}{m_h^2} \, \hat{t}_{1T}^{hh}
+ \frac{( {\Vec k_\perp} \cdot  {\Vec S}_\perp)
\, k_\perp^i}{m_N^2} \, \hat{t}_{1T}^{\perp \perp}
\nonumber \\
& & + \, \frac{({\Vec k}_\perp \cdot  {\Vec S}_\perp)
\, P_{h\perp}^i - ( {\Vec P}_{h\perp} \cdot {\Vec S}_\perp)
\, k_\perp^i }{m_N m_h} \, \hat{t}_{1T}^{\perp h}
\nonumber \\
& & + \, \frac{\epsilon_{\perp}^{ij} \, P_{h\perp j}}{m_h}
\, \hat{t}_1^h
+ \frac{\epsilon_{\perp}^{ij} \, k_{\perp j}}{m_N}
\, \hat{t}_1^{\perp}\,,
\label{tp-frf}
\end{eqnarray}
where ${\Vec k}_\perp$ is the quark transverse momentum and by the vector
product of two-dimensional vectors ${\bf a}$ and ${\bf b}$ we mean the
pseudo-scalar quantity $ {\bf a} \times  {\bf b} =
\epsilon^{i j} \, a_i b_j =  a   b  \, \sin (\phi_b - \phi_a)$.
All fracture functions depend on the scalar variables $x_B, k_\perp^2, \zeta,
P_{h\perp}^2$ and ${\Vec k}_\perp \cdot {\Vec P}_{h\perp}$.
For the production of a spinless hadron in the TFR one has~\cite{Anselmino:2011ss}:
\bq
& & \hspace{1.5cm}\frac{\D\sigma^{\ell(l,\lambda) + N(P_N,S) \to
\ell(l') + h(P)+X}}{\D x_B \, \D y \, \D \zeta \, \D^2{\Vec P}_{h\perp}\,
\D\phi_S} =
\frac{\alpha_{\rm em}^2}{Q^2 y} \, 
\\
&& \hspace{-0.5cm} \times \Bigg \{
\left [1 +(1-y)^2 \right ]
\, \sum_a e_a^2 \,
 \left  [\tilde{u}_1(x_B, \zeta, P_{h\perp}^2)
- S_T \, \frac{P_{h\perp}}{m_h}
\, \tilde{u}_{1T}^h(x_B, \zeta, P_{h\perp}^2) \, \sin (\phi_h - \phi_S)
\right ]
\nonumber \\
& &
\hspace{0.cm} + \,
\lambda \, y \, (2 - y )
\sum_a e_a^2 \, \left [
S_L \, \tilde{l}_{1L} (x_B, \zeta, P_{h\perp}^2)
+ \, S_T\, \frac{P_{h\perp}}{m_h}
\, \tilde{l}_{1T}^h (x_B, \zeta, P_{h\perp}^2) \, \cos (\phi_h - \phi_S)
\right ] \Bigg \}
\nonumber \,, \label{cross1}
\eq
where the ${\Vec k}_\perp$-integrated fracture functions are given as
\bq
&& \> \tilde{u}_1(x_B, \zeta, P_{h\perp}^2)
= \int \!\! \D^2 {\Vec k}_\perp \, \hat{u}_1 \,,
\quad
\tilde{u}_{1T}^h(x_B, \zeta, P_{h\perp}^2)
= \int \!\! \D^2 {\Vec k}_\perp \, \big( \hat{u}_{1T}^h
+ \frac{m_h}{m_N}
\frac{{\Vec k}_\perp \cdot {\Vec P}_{h\perp}}{P_{h\perp}^2}
\, \hat{u}_{1T}^{\perp} \big )\, ,
\label{intm2}
\nonumber \\
&& \quad \tilde{l}_{1L}(x_B, \zeta, P_{h\perp}^2)
= \int \!\! \D^2 {\Vec k}_\perp \, \hat{l}_{1L} \,,
\quad
\tilde{l}_{1T}^h(x_B, \zeta, P_{h\perp}^2)
= \int \!\! \D^2 {\Vec k}_\perp \, \big( \hat{l}_{1T}^h +
\frac{m_h}{m_N}
\frac{{\Vec k}_\perp \cdot {\Vec P}_{h\perp}}{P_{h\perp}^2}
\, \hat{l}_T^{\perp} \big )\,.
\label{intdeltam2}
\eq
We see that a single hadron production in the TFR of SIDIS does not provide access to all fracture functions. At LO the cross-section, with unpolarized leptons,
contains only the Sivers-like single spin azimuthal asymmetry.

\section{Double hadron leptoproduction (DSIDIS)}
In order to have access to all fracture functions one has to "measure" the
scattered quark transverse polarization, for example exploiting he Collins effect~\cite{Collins:1992kk} -- the azimuthal correlation of the fragmenting quark transverse polarization, ${\Vec s}'_T$, with the produced hadron transverse momentum, ${\Vec p}_\perp$:
\be
D(z,{\Vec p}_\perp) = D_1(z, p_\perp^2) + \frac{{\Vec p}_\perp \times
{\Vec s}'_T}{m_h}H_1^\perp(z, p_\perp^2)\, ,
\ee
where $s'_T=D_{nn}(y)\,s_T$ and $\phi_{s'}=\pi-\phi_s$ with
$D_{nn}(y)= [2(1-y)]/[1+(1-y)^2]\>$.

Let us consider a double hadron production process (DSIDIS)
\begin{equation}\label{dsidis}
l(\ell) + N(P) \to l(\ell') + h_1(P_1) + h_2(P_2) + X
\end{equation}
with (unpolarized) hadron 1 produced in the CFR ($x_{F1}>0$) and hadron 2 in the TFR ($x_{F2}<0)$, see Fig.~1.  For hadron $h_1$ we will use the ordinary scaled variable $z_1 = P_1^+/k'^+ \simeq P{\cdot}P_1/P{\cdot}q$ and its transverse
momentum ${\Vec P}_{1\perp}$ (with magnitude $P_{1\perp}$ and azimuthal angle
$\phi_1$) and for hadron $h_2$ the variables $\zeta_2 = P_2^-/P^- \simeq E_2/E$
and ${\Vec P}_{2\perp}$ ($P_{2\perp}$ and $\phi_2$).

\begin{figure}[h!]
\begin{center}
\label{fig:sidis-assoc}
  \includegraphics[height=.25\textheight]{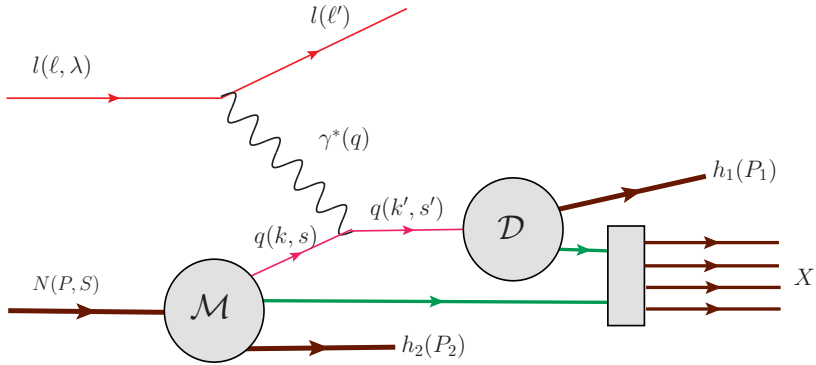}
  \caption{DSIDIS description in factorized approach at LO.}
\end{center}
\end{figure}

In this case the LO expression for the DSIDIS cross-section includes all fracture functions:
\bq \label{cs-2h}
&& \hspace{1.cm}
\frac{\D\sigma^{l(\ell,\lambda)+N(P,S) \to l(\ell')+h_1(P_1)+h_2(P_2)+X}}
{\D x \, \D y \, \D z_1 \, \D\zeta_2 \, \D^2 {\Vec P_{1\perp}} \,
\D^2 {\Vec P_{2\perp}} \, \D \phi_S} =
\frac{\alpha^2\,x_B}{Q^4 \, y}\left[ 1+(1-y)^2 \right] \times
\\ &&
\bigg(\mathcal{M}^{[\gamma^-]}_{h_2} \otimes D_{1q}^{h_1} + \lambda \,
D_{ll}(y)\, \mathcal{M}^{[\gamma^-\gamma_5]}_{h_2} \otimes D_{q}^{h_1}
+ \mathcal{M}^{[\I \, \sigma^{i -} \gamma_5]}_{h_2} \otimes
\frac{{\Vec p}_\perp \times  {\Vec s}'_T}{m_{h_1}}H_{1q}^{\perp h_1} \bigg) =
\nonumber \\
&& \hspace{-0.2cm}
\frac{\alpha^2\,x_B}{Q^4 \, y}\left[ 1+(1-y)^2 \right]
\left(\sigma_{UU} + S_\parallel \,\sigma_{UL} + S_\perp \,\sigma_{UT}+
\lambda \, D_{ll} \, \sigma_{LU} + \lambda \,S_\parallel D_{ll}\,\sigma_{LL}
+\lambda \, S_\perp D_{ll}\,\sigma_{LT} \right)\, ,
\nonumber
\eq
where $D_{ll}(y) = {y(2-y)}/{1+(1-y)^2}$ .

\section{DSIDIS cross-section integrated over ${\Vec P_{2\perp}}$}

If we integrate the fracture matrix over ${\Vec P}_{2\perp}$
we are left with eight $k_{\perp}$-dependent fracture functions:
\bq
\int \D^2 {\Vec P_{2\perp}} \, \mathcal{M}^{[\gamma^-]}
&=&
u_1 + \frac{{\Vec k_\perp} \times
\Vec S_{\perp}}{m_N} \, u_{1T}^{\perp} \>, \label{v1v2_tilde} \\
\int \D^2 {\Vec P_{2\perp}} \, \mathcal{M}^{[\gamma^- \gamma_5]}
&=&
S_{\parallel} \,  l_{1L}
+ \frac{{\Vec k_\perp} \cdot \Vec S_{\perp}}{m_N} \, l_{1T} \>,
\label{a1a2_tilde} \\
\int \D^2 {\Vec P_{2\perp}} \, \mathcal{M}^{[\I \, \sigma^{i -} \gamma_5]}
&=& S_{\perp}^i \,  t_{1T}
+ \frac{S_{\parallel} \, k_{\perp}^i}{m_N} \,
 t_{1L}^{\perp}
 + \frac{k_{\perp}^i ({\Vec k_\perp} \cdot \Vec S_{\perp})}{m_N^2}
\,   t_{1T}^{\perp}
+ \frac{\epsilon_{\perp}^{ij} k_{\perp j}}{m_N}
\,  t_1^{\perp}
\nonumber \\
&=& S_{\perp}^i \,  t_1
+ \frac{S_{\parallel} \, k_{\perp}^i}{m_N} \,
 t_{1L}^{\perp}
+ \frac{(k_{\perp}^i k_{\perp}^j - \frac{1}{2} {\Vec k_\perp}^2 \delta_{ij}) \,
S_{\perp}^j}{m_N^2} \,  t_{1T}^{\perp}
+ \frac{\epsilon_{\perp}^{ij} k_{\perp j}}{m_N}
\,  t_1^{\perp}\,,
\label{t1t2_tilde}
\eq
where $ t_{1} \equiv t_{1T}
+ ({\Vec k_\perp}^2/2 m_N^2) \,  t_{1T}^{\perp}$.
We have removed the hat to denote the ${{\Vec P}_{2\perp}}$--integrated
fracture functions, for example:
\be
t_1 (x_B, {\Vec k_\perp}^2, \zeta) =
 \int \D^2 {\Vec P_{2\perp}}
\left \{  \hat{t}_{1T} + \frac{{\Vec k_\perp}^2}{2m_N^2} \,
 \hat{t}_{1T}^{\perp\perp} + \frac{\Vec P_{2\perp}^2}{2m_2^2} \,
 \hat{t}_{1T}^{hh} \right \}.
\ee
The complete expression for other seven ${\Vec P_{2\perp}}$--integrated
fracture functions are presented in Ref.~\cite{Anselmino:2011bb}.

These ${\Vec P_{2\perp}}$--integrated fracture functions are perfectly
analogous to those describing single-hadron leptoproduction
in the CFR~\cite{Bacchetta:2006tn}, the correspondence being:
Fracture Functions $\Rightarrow$ Distribution Functions.
Thus we can use the procedure of Ref.~\cite{Bacchetta:2006tn} to obtain
the final expression of the cross section as
\bq
& &
 \frac{\D \sigma}{\D x_B \, \D y \, \D z_1 \, \D \zeta_2 \,
\D \phi_1 \, \D  P_{T1}^2 \, \D \phi_S} =
 \frac{\alpha_{\rm em}^2}{ x_B \, y \, Q^2} \left \{
\left (1 - y + \frac{y^2}{2} \right ) \, \mathcal{F}_{UU, T}
+ (1 - y) \, \cos 2 \phi_1 \, \mathcal{F}_{UU}^{\cos 2 \phi_1}  \right.
\nonumber \\
& & \hspace{1.3cm} + \, S_\parallel \,
 (1 - y) \, \sin 2 \phi_1 \, \mathcal{F}_{UL}^{\sin 2 \phi_1}
 + S_\parallel \, \lambda
\,  y \, \left (1 - \frac{y}{2} \right ) \, \mathcal{F}_{LL}
\nonumber \\
& & \hspace{1.3cm} + \, S_T \,
 \left (1 - y + \frac{y^2}{2} \right ) \, \sin (\phi_1 - \phi_S) \,
\mathcal{F}_{UT}^{\sin (\phi_1 - \phi_S)}
\nonumber \\
& & \hspace{1.3cm} + \, S_T \,  (1 -y) \, \sin (\phi_1 + \phi_S) \, \mathcal{F}_{UT}^{\sin (\phi_1 +
\phi_S)} + S_T \, (1 - y ) \, \sin (3 \phi_1 - \phi_s) \,
\mathcal{F}_{UT}^{\sin (3 \phi_1 - \phi_S)}
\nonumber \\
& & \hspace{1.3cm} + \left.  S_T \,  \lambda
\,  y \left (1 - \frac{y}{2} \right ) \, \cos (\phi_1 - \phi_S) \,
\mathcal{F}_{LT}^{\cos (\phi_1 - \phi_S)}
\right \}
\label{sidiscs_lt}
\eq
where the structure functions are given by the same convolutions as in~\cite{Bacchetta:2006tn} with the replacement of the TMDs with the
${\Vec P_{2\perp}}$--integrated fracture and fragmentation functions:
$f \to u, g \to l$ and $h \to t$.

\section{DSIDIS cross-section integrated over ${{\bf P}_{T1}}$}

If one integrates the DSIDIS cross-section over ${\Vec P_{1\perp}}$ and the
quark transverse momentum only one fragmentation function, $D_1$, survives,
which couples to the unpolarized and the longitudinally polarized
${\Vec k}_\perp$--integrated fracture functions:
\bq
 \int \D^2 {\Vec k_{\perp}} \, \mathcal{M}^{[\gamma^-]}
&=& \tilde{u}_1 (x_B, \zeta_2, P_{2\perp}^2)
+ \frac{{\Vec P_{2\perp}}\times \Vec S_T}{m_2}
\, \tilde{u}_{1T}^h (x_B, \zeta_2, P_{2\perp}^2),
\label{intkfrag1} \\
 \int \D^2 {\Vec k_{\perp}} \,
\mathcal{M}^{[\gamma^- \gamma_5]}
&=& S_\parallel \, \tilde{l}_{1L} (x_B, \zeta_2, P_{2\perp}^2)
+ \frac{{\Vec P_{2\perp}} \cdot \Vec S_T}{m_2}
\, \tilde{l}_{1T}^h (x_B, \zeta_2, P_{2\perp}^2),
\label{intkfrag2}
\eq
where the fracture functions with a tilde (which means integration
over the quark transverse momentum) are as in Eqs.~(\ref{intdeltam2}).
%

The final result for the cross section is~\cite{Anselmino:2011bb}
\bq
& &  \frac{\D \sigma}{\D x_B \, \D y \, \D z_1 \, \D \zeta_2 \,
\D \phi_2 \, \D  P_{2 \perp}^2 \, \D \phi_S} =
 \frac{\alpha_{\rm em}^2}{y \, Q^2} \,
\left \{ \left (1 - y + \frac{y^2}{2} \right ) \right.
\nonumber \\
& & \hspace{1cm}
\times \, \sum_a e_a^2 \,
 \left  [   \tilde{u}_1(x_B, \zeta_2, P_{2\perp}^2)
- S_T \, \frac{P_{2\perp}}{m_2}
\, \tilde{u}_{1T}^h (x_B, \zeta_2, P_{2\perp}^2) \, \sin (\phi_2 - \phi_S)
\right ]
\nonumber \\
& &  \hspace{1cm} + \,
\lambda \, y \, \left (1 - \frac{y}{2}\right )
\sum_a e_a^2 \,
 \left [ \STRUT
S_\parallel \, \tilde{l}_{1L} (x_B, \zeta_2, P_{2\perp}^2)
\right.
\nonumber \\
& & \hspace{1cm}
+ \, \left. \left.
S_T \, \frac{P_{2\perp}}{m_2}
\,  \tilde{l}_{1T}^h (x_B, \zeta_2, P_{2\perp}^2) \, \cos (\phi_2 - \phi_S)
\right ] \right \} D_1 (z) .
\label{crossintk}
\eq
As in the case of single-hadron production \cite{Anselmino:2011ss}, there
is a Sivers-type modulation $\sin (\phi_2 - \phi_S)$, but no Collins-type
effect.

\section{Examples of unintegrated cross-sections: beam spin asymmetry}

We show here explicit expressions only for $\sigma_{UU}$ and $\sigma_{LU}$\footnote{Expressions for other terms are available in~\cite{Kotzinian:DIS2011}.}
\bq\label{s_uu}
\sigma_{UU}  =  F_0^{{\hat u} \cdot D_1}
& - & D_{{nn}} \Bigg[\frac{P_{{1\perp}}^2 }{m_1 m_N}\, F_{{kp1}}^{{\hat t}^\perp \cdot H_1^\perp}\,{\cos}(2 \phi _1)
+ \frac{P_{{1\perp}} P_{{2\perp}} }{m_1 m_2}\, F_{{p1}}^{{\hat t}^h
\cdot H_1^\perp}\, {\cos}(\phi _1+\phi _2)\nonumber \\
& + & \left(\frac{P_{{2\perp}}^2 }{m_1 m_N}\, F_{{kp2}}^{{\hat t}^\perp
\cdot H_1^\perp} + \frac{P_{{2\perp}}^2 }{m_1 m_2}\, F_{{p2}}^{{\hat t^h}
\cdot H_1^\perp}\right)\, {\cos}(2 \phi _2)\Bigg].
\eq
\be
\sigma_{LU} = -\frac{ P_{{1\perp}} P_{{2\perp}}}{m_2 m_N}  F_{{k1}}^{{\hat l}^{\perp h}\cdot D_1} \, \sin(\phi _1-\phi _2)
\, ,
\ee
where the structure functions $F_{...}^{...}$ are specific convolutions~\cite{Kotzinian:DIS2011, abk3} of fracture and fragmentation functions depending on $x, z_1, \zeta_2, P_{1\perp}^2,  P_{2\perp}^2, {\Vec P}_{1\perp} \cdot
{\Vec P}_{2\perp}$.

We notice the presence of terms similar to the Boer-Mulders term appearing in the usual CFR of SIDIS. What is new in DSIDIS is the LO beam spin SSA, absent in the CFR of SIDIS.
We further notice that the DSIDIS structure functions may depend in principle on the relative azimuthal angle of the two hadrons, due to presence of the last term among their arguments: ${\Vec P}_{1\perp} \cdot {\Vec P}_{2\perp} =
P_{1\perp} P_{2\perp}\cos(\Delta \phi)$ with $\Delta \phi=\phi_1-\phi_2$.
This term arise from ${\Vec k}_\perp \cdot  {\Vec P}_\perp$ correlations in STMD fracture functions and can generate a long range correlation between hadrons produced in CFR and TFR. In practice it is convenient to chose as independent azimuthal angles $\Delta \phi$ and $\phi_2$.

Let us finally consider the beam spin asymmetry defined as
\be
A_{LU}(x, z_1, \zeta_2, P_{1\perp}^2,  P_{2\perp}^2, \Delta \phi) =
\frac{\int \D \phi_2 \, \sigma_{LU}}{\int \D \phi_2 \, \sigma_{UU}}=
\frac{-\frac{ P_{{1\perp}} P_{{2\perp}}}{m_2 m_N}  F_{{k1}}^{{\hat l}^{\perp h}\cdot D_1} \, \sin(\Delta \phi)}{F_0^{{\hat u} \cdot D_1}}\,\cdot
\ee
If one keeps only the linear terms of the corresponding fracture function
expansion in series of ${\Vec P}_{1\perp} \cdot {\Vec P}_{2\perp}$ one obtains
the following azimuthal dependence of DSIDIS beam spin asymmetry:
\be
A_{LU}(x, z_1, \zeta_2, P_{1\perp}^2, P_{2\perp}^2) =
a_1 \sin(\Delta \phi) + a_2 \sin(2\Delta \phi)
\ee
with the amplitudes $a_1,a_2$ independent of azimuthal angles.

We stress that the ideal opportunities to test the predictions of the present
approach to DSIDIS, would be the future JLab 12 upgrade, in progress, and the
EIC facilities, in the planning phase.

\end{document}